# Toward perfect reads: self-correction of short reads via mapping on de Bruijn graphs

Antoine Limasset [1], Jean-François Flot [1,*], and Pierre Peterlongo [2,*]

[1] Université libre de Bruxelles, Evolutionary Biology & Ecology, C.P. 160/12, Avenue F.D. Roosevelt 50, 1050 Bruxelles, Belgium
[2] Univ Rennes, Inria, CNRS, IRISA, F-35000 Rennes, France

*Co-last authors with equal contributions.

Associate Editor: D

## Abstract

**Motivations** Short-read accuracy is important for downstream analyses such as genome assembly and hybrid long-read correction. Despite much work on short-read correction, present-day correctors either do not scale well on large data sets or consider reads as mere suites of k-mers, without taking into account their full-length read information.

**Results** We propose a new method to correct short reads using de Bruijn graphs, and implement it as a tool called Bcool. As a first step, Bcool constructs a compacted de Bruijn graph from the reads. This graph is filtered on the basis of $k$-mer abundance then of unitig abundance, thereby removing from most sequencing errors. The cleaned graph is then used as a reference on which the reads are mapped to correct them. We show that this approach yields more accurate reads than $k$-mer-spectrum correctors while being scalable to human-size genomic datasets and beyond.

**Availability and Implementation** The implementation is open source and available at http://github.com/Malfoy/BCOOL under the Affero GPL license.

**Contact** Antoine Limasset `antoine.limasset@gmail.com` & Jean-François Flot `jflot@ulb.ac.be` & Pierre Peterlongo `pierre.peterlongo@inria.fr`

## 1 Introduction

### 1.1 Why correct reads?

Genome sequencing is a fast-changing field. Two decades have seen three generations of sequencing technologies: Sanger electropherograms (a.k.a. first-generation sequencing), short reads from second-generation sequencing (SGS) and long, error-prone reads from third-generation sequencing (TGS). Albeit powerful, these technologies all come with stochastic errors, and for some, non-stochastic ones. Stochastic errors are usually corrected using a consensus approach leveraging the high coverage depth available in most genomic projects, whereas non-stochastic errors can be eliminated by "polishing" the sequences using reads obtained from a different sequencing technique.

In *de novo* assemblers following the overlap-layout-consensus (OLC) paradigm, the stochastic errors present in the reads are corrected during the consensus step toward the end of the process. By contrast, the de Bruijn graph (DBG) assembly paradigm does not include *per se* any error correction step (although error correction is an optional preliminary step proposed by many DBG assemblers); rather, DBG assemblers attempt to filter out erroneous $k$-mers by considering only $k$-mers present at least a minimal number of times in the reads to be assembled. In the case of DBG assemblers, lowering the error rate in the reads to be assembled makes it possible to use a greater $k$-mer size, paving the way for a more contiguous assembly. This being said, even OLC assemblers may benefit from a preliminary error correction step as it allows more stringent alignment parameters to be used, thereby improving the speed of the process and reducing the amount of spurious overlaps detected between reads.

Beyond *de novo* assembly, other applications that require mapping, such as SNP calling, genotyping or taxonomic assignation, may also benefit from a preliminary error-correction step aimed at increasing the signal/noise ratio and/or reducing the computational cost of detecting errors *a posteriori* [1, 2].

### 1.2 On the use of short reads as long reads are rising

Although long-reads technologies from third-generation sequencing (TGS), marketed by Pacific Biosciences (PacBio) and Oxford Nanopore Techonologies (ONT), are on the rise and may surpass short reads for many purposes such as genome assembly (as TGS reads are order of magnitude longer and are therefore less sensible to repeats [3]), we have reasons to think that SGS will still be broadly used in the next decade.

This because SGS reads remain considerably cheaper, and their accurate sequences make them highly valuable. Besides, recent methods are able to deliver long-distance information based on short-read sequencing. Chromosome conformation capture [4] provides pairs of reads that have a high probability to originate from the same chromosome and Chromium 10X [5] uses a droplet mechanism to ensure that a pack of reads come from a single DNA fragment up to hundred thousands base pairs. Both of these techniques have been shown to produce assembly continuity comparable to TGS assembly [6, 7, 8]. SGS reads are also used jointly with long reads to compensate the latter's high error rate (and systematic homopolymer errors in the case of ONT reads [9]) in a cost-efficient way [10]. These different applications make it worth investing effort in improving short-read correctors beyond the current state of the art, in the hope that near-perfect reads will positively impact all downstream analyses that require accurate sequences.

### 1.3 $k$-mer-spectrum techniques

$k$-mer-spectrum techniques are conceptually the simplest correction method and remain broadly used. The underlying intuition is that true genomic $k$-mers will be seen many times in the read set, whereas erroneous $k$-mers originating from sequencing errors will be comparatively much rarer. The first step to correct reads using this approach is therefore to choose an abundance threshold (above which $k$-mers are called "solid" and below which they are called "weak" [11]). $k$-mer-spectrum correctors aim to detect all weak $k$-mers in the reads and correct them by turning them into solid ones.

One of the best $k$-mer-spectrum correctors available to date [12] is Musket [13]; however, its memory consumption is high on large genomes because of its indexing structure. Another tool, Bloocoo [14], achieves a comparatively lower memory footprint, even on genomes comprising billions of bases (such as the human one), by using a Bloom filter to index $k$-mers. Lighter [15] also uses Bloom filters but bypasses the $k$-mer counting phase by only looking at a subset of the $k$-mers in a given data set, therefore achieving greater speed. One problem with these approaches above is they they apply greedy strategies and never revert their decisions, which can lead to suboptimal corrections on complex data sets. A less greedy approach is implemented in the BFC [16] corrector, which attempts to correct each read as a whole by finding the minimal number of substitutions required for a read to be entirely covered by solid $k$-mers.

### 1.4 Other read correction techniques

Other correction techniques rely either on suffix arrays (allowing the use of substrings of various sizes instead of only fixed $k$-length words) [17] or on multiple-read alignments [18]. Despite their methodological appeal, these techniques are resource-expensive and do not scale well on large data sets. Moreover, benchmarks suggest that they perform significantly worse than state-of-the-art $k$-mer-spectrum correctors [19]. Yet another approach for correcting reads, pioneered by LoRDEC [20] then by LoRMA [21], is to use de Bruijn graphs instead of strings as a basis for correction. In the LoRDEC approach, this de Bruijn graph is built from highly accurate short reads and used to correct long reads. In LoRMA, the de Bruijn graph is built from the very same long reads that the program is attempting to correct. Both programs were engineered to handle long reads and it is surprising that such DBG-based approach was never applied till now to correct highly accurate SGS short reads such as those produced by Illumina sequencers.

Here, we implement a DBG-based corrector geared towards Illumina reads, which are characterized by errors that are solely substitutions (no indels) and affect less than one percent of the output bases. We then that this corrector, dubbed Bcool, vastly outperforms state-of-the-art $k$-mer-spectrum correctors while being both scalable and resource-efficient.

## 2 Method

The intuition behind $k$-mer-spectrum correction is that $k$-mers, once filtered according to their abundance, represent a reference that can be used to correct the reads. The idea that a de Bruijn graph provides a better reference than a $k$-mer set might be surprising at first glance since a de Bruijn graph is equivalent to its set of $k$-mers. However, the novelty of our approach is that we build a compacted DBG, that is, a DBG in which non-branching paths are turned into unitigs. This results in a graph that is nearly error-free, allowing a DBG-based corrector to vastly surpass $k$-mer-spectrum ones.

After briefly reviewing several limitations of $k$-mer-spectrum correction (Section 2.1), we detail how our proposed approach manages to tackle these issues while describing some key parts of our workflow (Section 2.2).

### 2.1 $k$-mer spectrum limitations

In this section we identify four sources of miss-correction in $k$-mer spectrum approaches, as represented in Figure 1. As mentioned previously, $k$-mer-spectrum correctors infer a set of solid $k$-mers that are used for correcting reads. Erroneous $k$-mers (i.e., $k$-mers containing at least one sequencing error) can be filtered out by keeping only $k$-mers above a given abundance threshold called the solidity threshold. $k$-mers whose abundance is higher or equal (respectively lower) to this threshold are called solid (respectively weak).

*(1) Weak genomic $k$-mers.* Depending on the solidity threshold chosen, random variations in sequencing depth may cause some genomic $k$-mers to fall below the threshold and be erroneously filtered out. This kind of $k$-mer creates a situation such as the one represented in Figure 1.1, where an isolated weak $k$-mer is flagged as a putative error on a read that is actually correct. Since at least $k$ successive weak $k$-mers are expected to be seen when there is a sequencing error (Figure 2), $k$-mer spectrum-based correctors normally consider isolated weak $k$-mers as likely to be simply missed genomic $k$-mers and do not attempt to correct them.

*(2) Solid erroneous $k$-mers.* Conversely, setting the solidity threshold too low may lead to the inclusion of some erroneous $k$-mers in the trusted set of solid $k$-mers (Figure 1.2). As a result, the errors in the reads harboring such $k$-mers are not corrected and may also propagate to other reads if these $k$-mers are used for correction.

*(3) Errors covered by solid $k$-mers* If a sequencing error in a read is covered partly or entirely by genomic $k$-mers originating from other regions of the genome, the error may not be detected and it may be difficult to correct as the remaining isolated weak $k$-mers will be hard to distinguish from the pattern in point 1 above. Such situations are likely to occur in repeated or quasi-repeated genome regions, leading to complex situations where genomic sequences may have several contexts.

*(4) Nearby errors* Accurate correction is also complex when multiple errors occur nearby each other (i.e., less than $k$ bases apart). In such situations, the number of errors and their positions cannot be easily estimated, and $k$-mer-spectrum correctors have to perform a very large number of queries to correct them. Musket uses an aggressive greedy heuristic that tries to replace the first weak $k$-mer encountered by a solid one then checks whether the next $k$-mers became solid as a result. But, as shown in Figure 1.4, this heuristic is inefficient if the $k$-mers that follow contain other sequencing errors.

*$k$-mer size* All the issues highlighted above boil down to a central problem when using $k$-mer-spectrum correctors: the size of $k$. If a too large $k$-mer size is used, most $k$-mers contain at least one sequencing error and many of them actually contain several errors (Figure 2). In those cases

| | 1) Weak genomic *k*-mers | 2) Solid erroneous *k*-mers | 3) Errors covered by solid *k*-mers | 4) Nearby errors |
|---|---|---|---|---|
| reference | GCTGATCGCTAGTT | GCTGATCGCTAGTT | GCTGATCGCTAGTT....AGCTCTTT | GCTGATCGCTAGTT |
| set of solid *k*-mers (erroneous bases) | GCTG CTGA TGAT GATC ATCG TCGC CGCT GCTA CTAG TAGT AGTT | GCTG CTGA TGAT GATC ATCG TCGC CGCT GCTA CTAG TAGT AGTT CGTT TCGT GCTC CTCG | GCTG CTGA TGAT GATC ATCG TCGC CGCT GCTA CTAG TAGT AGTT AGCT GCTC CTCT TCTT CTTT | GCTG CTGA TGAT GATC ATCG TCGC CGCT GCTA CTAG TAGT AGTT |
| example of read (weak *k*-mers, solid *k*-mers) | GCTGATCGCTAGTT | GCTGATCGCTCGTT | GCTGATCGCTCGTT | GCTGAGCCCTAGTT |
| | ATCG is genomic but weak | GCTC is erroneous but solid | GCTC is solid as it exists somewhere else in the genome | two nearby errors in read result in six weak *k*-mers |

**Fig. 1.** Four issues with *k*-mer-spectrum methods. 1) Genomic *k*-mers may be appear weak because of their low abundance. 2) Erroneous *k*-mers may appear solid because of their high abundance. 3) Sequencing errors may be validated by genomic *k*-mers originating from other parts of the genome. 4) Multiple errors may occur on a *k*-mer, resulting in a large weak region.

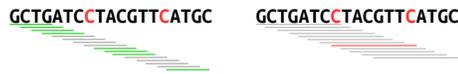

**Fig. 2.** Patterns of weak *k*-mers resulting from a sequencing error, depending on *k*. Sequencing errors in the read are pictured in red. Solid and weak *k*-mers are represented respectively with green and grey lines, with $k = 4$ (left) and $k = 8$ (right). The *k*-mer pictured in red contain two errors. A sequencing error usually impacts *k* *k*-mers and creates a weak region of size $2 * k - 1$ (left). Choosing a large *k* results in most *k*-mers of a read being weak as they contain one or several errors (right).

*k*-mer-spectrum correctors may be unable to locate errors and to perform correction as they rely on genomic *k*-mers to find possible candidates to replace weak ones. On the contrary, if *k* is too small, most *k*-mers are solid and almost no correction is performed. As *k*-mer-spectrum correctors are usually geared towards correcting SGS reads, they use a *k*-mer size around 31 that is well suited for the error rate of Illumina data. Choosing a larger *k* results in sub-optimal correction or even in a failure of the program to run (see Supplementary materials, Table 5 for more information). This limitation may be a problem when addressing large and repeat-rich genomes, as a large number of *k*-mers are repeated in various contexts throughout the genomes and large *k*-mers are required in order to distinguish them.

## 2.2 DBG-based reads correction

In this section we describe our proposal, called Bcool (which stands for "de Bruijn graph-based read correction by alignment"). The basic idea is to construct a DBG from the read set, to clean it, and then to align the reads on the DBG. Reads that map on the DBG with less than a threshold number of mismatches are corrected using the graph sequence, which is supposed to be almost error-free. An important Bcool feature is that the graph is constructed by filtering out low-coverage *k*-mers and additionally by discarding low-coverage unitigs (see Section 2.2.2), yielding a reference graph with a very low amount of erroneous *k*-mers.

We present in Figure 3 some simple examples illustrating how our DBG-based read correction handles the problems highlighted in Figure 1. Our proposal differs from *k*-mer-spectrum techniques in that our reference is a compacted DBG [22] instead of a set of *k*-mers, and that we map the reads onto the graph instead of looking at all *k*-mers contained in the reads.

Bcool's workflow is depicted in Figure 4. Each of its components is either an independent tool already published or an independent module that could be reused in other frameworks. We describe below the different key steps of the workflow.

### 2.2.1 DBG construction

In Bcool, the DBG is constructed using Bcalm2 [24], a resource-efficient method to build a compacted DBG. In a compacted DBG, nodes are not composed of single *k*-mers but of unitigs (i.e., maximal simple paths of the DBG) of lengths larger or equal to *k*. As explained in more detail in the next section, Bcool uses a seed-and-extend mapping strategy with seeds of length inferior or equal to the *k* parameter used for graph construction. This allows Bcool to correct reads that do not contain any solid *k*-mer, as long as these reads contain at least one error-free seed and align on the graph. Thus, Bcool can use a large *k*-mer size and is therefore less affected by genomic repeats than *k*-mer-spectrum correctors. Moreover, with a large *k* value, most sequencing errors generate a tip in the graph rather than a bubble. Bcool takes advantage of this by performing a graph-correction step aimed at removing tips.

### 2.2.2 DBG clean-up

A specificity of this work stands is the way we clean up the reference DBG before using it to correct reads. In this context we propose and use the notion of *unitig abundance*, defined as the the mean abundance of all its constituent *k*-mers.

The DBG is initially constructed with a very low abundance threshold (by default 2, i.e. *k*-mers that occur only once are considered as probable errors and discarded). This very low threshold value decreases the probability of missing genomic *k*-mers, but as a consequence, many erroneous *k*-mers are not filtered out. A first step to remove those is to remove short dead-ends (a.k.a 'tips'). Formally, we define a tip as a unitig of length inferior to $2 * (k - 1)$ that has no successor at one of its extremities. Such dead-ends mainly result from sequencing errors occurring on the first or last *k* nucleotides of a read. By contrast, errors located at least *k* nucleotides away from the start or the end of a read form bubble-like patterns, and such errors are detected and filtered based on unitig coverage. In this second step, we tackle remaining erroneous *k*-mers by taking a look at unitig abundance. We choose a unitig abundance threshold (higher than the *k*-mer abundance threshold used previously) and when an unitig has an abundance lower than this threshold, we discard it completely. Intuitively, averaging the abundance across each unitig makes it possible to 'rescue' genomic *k*-mers with low abundance (that tend to be lost by *k*-mer-spectrum techniques, see Figure 1.1) by detecting that they belong to high-abundance unitigs, and these genomic *k*-mers can then be used for correction. On the other hand, erroneous *k*-mers are likely to belong to low-abundance unitigs that are filtered out. The unitig threshold can be user-specified or can be inferred by looking at the unitig abundance distribution and choosing the first local minimum. These cleaning steps are applied several time in an iterative manner for handling complex scenarios where error patterns are nested. Together, those two approaches allow us to address the issues depicted in Figure 1.1 and Figure 1.2. Compared with a strategy based only on *k*-mer abundance, our approach keeps more low-abundance genomic *k*-mers while removing more erroneous *k*-mers. As shown in Figure 5 using simulated data, the filtering strategy used by Bcool produces markedly less false negatives (FN) and false positives (FP) than the sole *k*-mer-abundance threshold used by *k*-mer-spectrum correctors, resulting in a better set of *k*-mers. Detailed evaluation of the tipping and unitig-filtering strategies is provided in the Supplementary Materials (Section 10).

| | 1) Weak genomic *k*-mers | 2) Solid erroneous *k*-mers | 3) Errors covered by solid *k*-mers | 4) Nearby errors |
|---|---|---|---|---|
| reference | GCTGATCGCTAGTT | GCTGATCGCTAGTT | GCTGATCGCTAGTT…AGCTCTTT | GCTGATCGCTAGTT |
| example of read<br>erroneous bases | GCTGATCGCTAGTT | GCTGATCGCTCGTT | GCTGATCGCTCGTT | GCTGAGCCTAGTT |
| Bcool solution<br>• mismatches | GCTGATCGCTAGTT | pruned tip<br>CTCGTT<br>GCTGATCGCTA<br>CTAGTTT | GCTAGTT best mapping<br>GCTGATCGCT<br>GCTCTTT<br>…AGCT | GCTGATCGCTAGTT |

**Fig. 3.** How Bcool handles the problems highlighted in Figure 1. Blue half-arrows represent the paths of the graph on which given read maps. 1) By using a very low *k*-mer abundance threshold coupled with a unitig abundance threshold, Bcool retains low-abundance *k*-mers and manages to correct the reads that contain them. 2) Bcool detects the tip pattern produced by solid erroneous *k*-mers and is therefore able to discard them. Other erroneous *k*-mers are detected at the unitig filtering step. 3) By considering mappings globally, Bcool chooses the best path for each read, i.e., the one on which it maps with the smallest number of mismatches. 4) Bcool uses unitigs instead of *k*-mers to correct reads and is therefore able to deal with reads that contain several nearby errors.

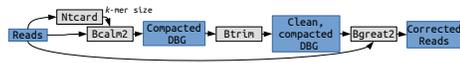

**Fig. 4.** Bcool workflow. The blue boxes are FASTA files and the grey boxes represent the tools that process or generate them. Ntcard [23] is used to select the best-suited *k*-mer size. A compacted DBG is then constructed using Bcalm2 [24]. The Btrim [25] module cleans the graph, and the reads are finally mapped back on the de Bruijn graph using Bgreat2 [26].

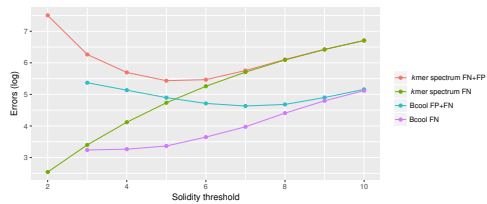

**Fig. 5.** Impact of the solidity threshold on the number of false positives (FP, erroneous *k*-mers that are retained in the *k*-mer set) and false negatives (FN, genomic *k*-mers that are discarded) with $k = 63$ on a 50X coverage of 150-bp reads simulated from the C. elegans reference genome. For *k*-mer-spectrum techniques, the solidity threshold applies to *k*-mers, while for Bcool it applies to the unitigs constructed from non-unique *k*-mers.

### 2.2.3 Read mapping

In contrast to *k*-mer spectrum-based techniques, Bcool uses an explicit representation of the DBG. Although in its current implementation this entails a higher memory usage and computational cost than *k*-mer-spectrum correctors, doing so provides an efficient way to fix the issues depicted in Figure 1.3 and Figure 1.4. Each read is aligned in full length on the graph, and the correction is made on the basis of the most parsimonious path on which the read maps in the graph. For mapping reads on the DBG, we use an improved, yet unpublished, version of Bgreat [27] called "Bgreat2". The main improvements are that Bgreat2 has no third-party dependencies (in contrary to the published version) and that it outputs optimal alignments among the candidates. The alignment procedure uses a classical seed-and-extend process. The extend phase allows only substitutions (at most 10 by default). Using the graph, the extend phase maps a read on several potential paths, in a breadth-first approach. Among all valid alignments, only those minimizing the number of mismatches are considered. If several different optimal alignments exist, by default the read is not mapped. This choice can optionally be modified to output one of the optimal mappings. In order to keep memory usage low, Bgreat2 uses a minimal perfect hash function [28] for indexing seeds. Moreover, it is possible to sub-sample seeds for indexation. For instance, by indexing only one out of ten seeds, we were able to run Bgreat2 on a human data set using around 30GB of RAM at the price of only a slight decrease in correction efficiency (see Table 1).

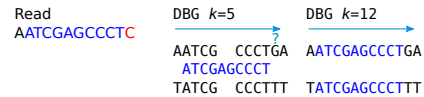

**Fig. 6.** Using a large *k*-mer size simplifies the graph by lowering the amount of unsolved repeats. In this example a repeat of size 10 is present in the genome in different contexts. With $k = 5$, we are not able to correct the last nucleotide of the read represented by a blue arrow. But with $k = 12$ we have determined the context of the repeat and know that only two possible paths exist. This way we are able to safely correct the read.

### 2.2.4 *k*-mer size selection

We use the highest possible *k* value. This has the effect to resolve repeats smaller than *k*, thereby improving the correction as shown on Figure 6. However, choosing a *k* value too large would yield a fragmented graph. Therefore, we implemented an automated tool, somewhat similar to *k*-merGenie [29], that uses ntCard to estimate the *k*-mer spectrum of the data set for several values of *k*. Our approach detects the first local minimum for each *k*-mer spectrum then selects the highest *k* value for which this minimum is above the unitig threshold. This way, we expect to keep most genomic *k*-mers that are more abundant than the unitig threshold. This approach is more conservative and simpler than the one implemented in *k*-merGenie, which attempts to fit the *k*-mer spectrum on a haploid or diploid model with the aim of finding the *k* value most suitable for assembling the reads.

| Corrector | RAM used (GB) | Wall-clock time (H:min) | CPU time (H:min) | Correction ratio |
|---|---|---|---|---|
| *C. elegans* | | | | |
| Bloocoo | 5.462 | 0:07 | 2:02 | 30.28 |
| Lighter | **0.627** | **0:06** | **1:40** | 31.16 |
| Musket | 24.755 | 0:56 | 16:44 | 90.33 |
| BFC | 8.390 | 0:13 | 4:29 | 14.58 |
| Bcool | 12.449 | 0:21 | 4:25 | **183.53** |
| Human | | | | |
| Bloocoo | **10.500** | 9:31 | 90:10 | 6.79 |
| Lighter | 14.121 | **4:22** | **60:06** | 5.65 |
| Bcool | 178.885 | 19:10 | 265:49 | **77.94** |
| Bcool i10 | 29.960 | 27:17 | 445:57 | 76.83 |

Table 1. Performance comparison on C. elegans and human simulated 250-bp reads with 1% error rate and 100X coverage using 20 cores. Bcool i10 indexed only one out of every ten seeds to reduce memory usage. BFC and Musket were not able to correct the full human data sets.

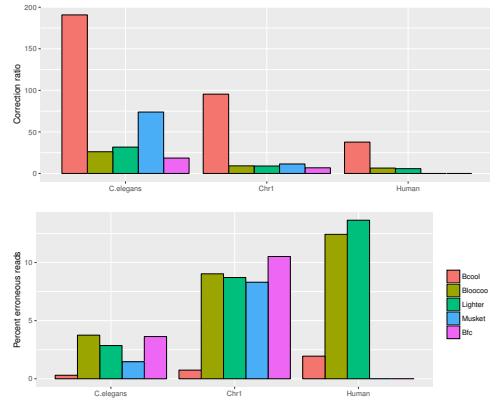

**Fig. 7.** Correction ratios (top) and percentage of erroneous reads after correction (bottom) for different correctors on our three simulated haploid data sets. We simulated 100x of 150bp reads with a 1% error rate. BFC and Musket ran out of memory on all full human data sets.

## 3 Results

We present results based on simulated data sets as well as on real ones. Simulations make it possible to precisely evaluate correction metrics (Section 3.2) and to assess their impact on downstream assembly (Section 3.3). Correction evaluation was performed using simulated reads from several reference genomes: *C. elegans*, the human chromosome 1, and the whole human genome. By contrast, the results presented in Section 3.4 aim to validate our approach using real data. All experiments were performed on a 20-core cluster with 250GB of RAM. Our results are compared with those obtained using several state-of-the-art short-read correctors: Bloocoo [14], Musket [13], BFC [16], and Lighter [15]. We tried to include LoRDEC in our benchmark (since this long-read corrector rests on a principle similar to Bcool) but we finally let it out as we did not manage to obtain results on par with programs designed for correcting short reads. In what follows, False Negatives ($FN$) stand for non-corrected errors, whereas False Positives ($FP$) are erroneous corrections and True Positives ($TP$) are errors that were correctly corrected. The *correction ratio* is defined as $= \frac{TP+FN}{FN+FP}$; it is the ratio of the number of errors prior to correction ($TP + FN$) vs. after correction ($FN + FP$). The higher the correction ratio, the more efficient the tool.

### 3.1 Performance benchmark

Before presenting qualitative results, we first compare the performance of the correctors included in our benchmark. We evaluated the resources used by the different correctors on data sets simulated from the *C. elegans* and human genomes. We report here the memory used, the wall-clock time and the CPU time reported by the Unix *time* command. Our results, presented in Table 1, show that that Bcool has a higher RAM footprint and is slower than the other tools we tested, except Musket. This is due to Bcool's explicit graph representation and its indexation scheme. However, Bcool scales well with genome size, as shown in Table 1. Moreover, it is possible to reduce the memory footprint by sub-sampling during indexation the seeds used for read mapping. This results in a greatly reduced memory footprint at the price of a slight decrease in correction performance. In the human experiment, graph creation took ≈8h30 and read mapping took ≈10h. As discussed below, there is clearly room for performance improvements both during the graph construction phase and the read mapping phase.

### 3.2 Correction ratios on simulated data sets

Our results on simulated haploid data are presented in Figure 7. They show that Bcool obtained a correction ratio one order of magnitude above the other tested tools. Note that, as shown in Supplementary materials (Sections 6 to 9) we tested several other conditions (longer reads, lower coverage, lower error rate), all leading to the same conclusion. In each of our experiments, Bcool had a better correction ratio together with a better precision and recall. The correction precision is critical, as a corrector should not introduce new errors. Our experiments showed a good precision for all the tools we tested (even on a human genome), with a net advantage for Bcool though. For instance, our human-genome experiment with 100x coverage of 150bp reads yielded a precision of 95.33% for Bloocoo, 96.74% for Lighter and 99.61% for Bcool.

*Diploid correction.* To assess the impact of heterozygosity, we tested these correctors on simulated human diploid genomes. To obtain a realistic distribution of SNPs and genotypes, we used SNPs predicted from real human individuals and included them in the reference genome. Our results show that, on these data sets, the result quality of all tested correctors remains almost identical to those obtained on haploid simulations. The details and results of this experiment are presented in the Supplementary Material.

### 3.3 Impact of the correction on assembly

In order to evaluate the impact of the correction on assembly, we ran the Minia [30] assembler on uncorrected reads and on read sets corrected with each of the correctors included in our benchmark. For each assembly, we tested several $k$-values (the main parameter of Minia), from $k = 21$ to $k = 141$ with a step of 10. For each corrector, only the best result is presented here. These results, presented in Figure 8, show that the N50 assembly metric is better on data corrected by Bcool. This can be explained by the fact that with a better read correction, a higher $k$ value can be selected, leading to a more contiguous assembly. As an example, for the *C. elegans* genome with 250-bp reads the best $k$-mer size was 91 for the raw reads, 131 for reads corrected using Lighter or Musket, 141 for reads corrected using Bloocoo and 171 for reads corrected using Bcool.

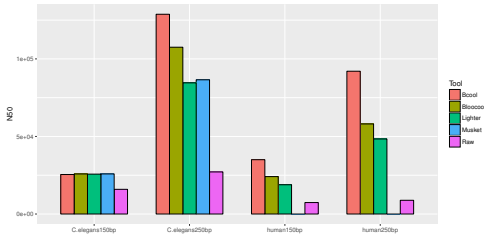

**Fig. 8.** Comparison of the N50s of the best assemblies obtained from the corrected simulated reads using Minia [30].

### 3.4 Real data sets

In this section we evaluate the impact of the correction on assembly continuity using several real data sets: a *C. elegans* Illumina HiSeq 2500 run with 79.8 millions reads of length 150 bp amounting to 12 Gbp (DRR050374); and a *A. thaliana* Illumina MiSeq run with 33.6 millions reads of length 250 bp amounting to 4.4 Gbp (ERR2173372). For this benchmark we used the string graph assembler fermi [31] given its efficiency and robustness. Assembly reports provided by Quast [32] are presented in Tables 2 and 3 using contigs larger than 1000 nucleotides. We see that on both datasets that Bcool obtains the most contiguous assembly in view of both N50 and N75 statistics.

| Corrector | N50 | NGA50 | N75 | NGA75 | Nb contigs |
|---|---|---|---|---|---|
| Bcool | **23,021** | **20,696** | **10,880** | **8,552** | 8,373 |
| BFC | 19,524 | 17,872 | 9,490 | 7,642 | 9,235 |
| Bloocoo | 20,968 | 18,960 | 10,105 | 7,993 | 8,836 |
| Lighter | 21,435 | 19,269 | 10,254 | 8,122 | 8,714 |
| Musket | 22,545 | 20,291 | 10,788 | 8,545 | **8,368** |

Table 2. Results of assembling the Illumina dataset DRR05374 (C. elegans) using fermi following various read correctors.

| Corrector | N50 | NGA50 | N75 | NGA75 | Nb contigs |
|---|---|---|---|---|---|
| Bcool | **34,206** | **24,466** | **16,690** | **9,301** | 8,419 |
| BFC | 29,002 | 21,634 | 15,363 | 9,254 | **8,376** |
| Bloocoo | 27,737 | 20,751 | 14,281 | 8,783 | 8,783 |
| Lighter | 25,274 | 19,325 | 13,194 | 8,277 | 9,420 |
| Musket | 27,348 | 20,453 | 14,021 | 8,628 | 9,139 |

Table 3. Results of assembling the Illumina dataset ERR2173372 (A. thaliana) using fermi following various read correctors.

## 4 Perspective

### 4.1 Perspectives regarding short reads

We have shown how to construct and clean a reference graph that can be used to efficiently correct sequencing errors. This approach is not to be compared with multiple-$k$ assembly as here we only apply a conservative correction to the graph without trying to remove variants nor to apply heuristics to improve the graph continuity: only $k$-mers that are very likely to be erroneous are removed in this process. Such conservative modifications on an intermediate graph used as a reference appears to us a promising approach to better exploit the high accuracy of short reads. The use of a high $k$-mer size is critical to address the correction problem on large, repeat-rich genomes, and the impossibility of $k$-mer-spectrum correctors to use a large $k$-mer size is a major limitations of such approaches. In contrast, our DBG-based solution uses a large $k$-mer size and therefore yields a more efficient correction on large, repeat-rich genomes. The resulting error-corrected reads are nearly perfect and can be assembled using an overlap-graph algorithm or may be used for other applications, such as variant calling.

Several propositions can be made to further improve Bcool. The read mapping step could make use of the quality values available in FASTQ files or provide other types of correction, such as read trimming. Adding the capacity to detect and correct indels during the mapping step could allow Bcool to correct other types of sequencing data, such as Ion torrent or PacBio CCS reads. The performance of the pipeline could also be globally improved. The de Bruijn graph construction could implement techniques similar to the sub-sampling used by Lighter in order to reduce its reliance on disk and therefore improve its running time. Besides, our mapping method is still naive, and implementing efficient heuristics such as the ones used by BWA [33] and Bowtie2 [34] could greatly improve the throughput of Bcool without decreasing the quality of the alignment. Our $k$-mer-spectrum analysis could also be improved to choose a more accurate $k$-mer size and abundance filter at both $k$-mer and unitig level on real, on haploid or heterozygous data. From a more theoretical viewpoint, a study of whether using successively multiple $k$-mer sizes provides an even better correction (albeit at the price of a longer running time) would be an interesting perspective. Last but not least, another possible development could look into applications to data sets with highly heterogeneous coverage, as observed in single-cell, transcriptome or metagenome sequencing data.

### 4.2 Perspectives regarding long reads

Surprisingly, the idea of aligning reads on a de Bruijn graph was applied to long, noisy reads before short, accurate ones. The efficiency of LoRDEC [20] and Bcool, respectively on long and short reads, suggests that such DBG-based correction is a general approach that can be applied to various kind of data sets. Short reads can also be used in conjunction with long reads, as for correcting systematic errors such as ONT homopolymers [9].

Using nearly perfect short reads as those corrected by Bcool may also improve long read correction. To test it, we simulated both short and long reads from the *C.elegans* reference genome and compared the amount of errors still present in the long reads after LoRDEC hybrid correction by mapping them on the reference with BWA. A coverage of 100X of short reads of 150 bp were simulated along with long reads with 12% error rate using Pbsim [35]. Applying LoRDEC using non-corrected short reads lead to a 3.04% error rate on corrected long reads. When the short reads were first corrected using Bcool, the error rate on the corrected long reads fell to 2.33% (a 30.5% improvement). Additional work will be required to explore this idea further.

Last but not least, in the current context of decreasing error rates for long reads, we may soon reach a point at which $k$-mer or DBG-based technique will manage to perform efficient *de novo* reference-based correction using long reads alone. LORMA [21] is a first such attempt at using de Bruijn graph created from long reads to perform pure correction in an iterative manner. This suggests that de Bruijn graphs still have a bright future in bioinformatics.

## 5 Acknowledgment

We thank Romain Feron and Camille Marchet for their support and implication in this project. Antoine Limasset's postdoctoral position is funded by the Fédération Wallonie-Bruxelles via a "Action de Recherche Concertée" (ARC) grant to Jean-François Flot. Computational resources were provided by the Consortium des Équipements de Calcul Intensif (CÉCI), funded by the Fonds de la Recherche Scientifique de Belgique (F.R.S.-FNRS) under Grant No. 2.5020.11, as well as by GenOuest platform (genouest.org).

## Supplementary materials

Sections 6 to 9 provide clarifications and additional results to those shown in Section 3.2 (longer reads, lower coverage, lower error rate, distinct $k$ values, diploid simulations). For all results except those presented in Table 5 we used the default $k$ value of each tool. All simulations (except in Section 8) are directly made from reference genomes and do not contain diploid variations.

Section 10 provides additional information on the tipping and unitig-filtering strategies.

For all presented results, the sensitivity is given by $\frac{TP}{TP+FN}$ and the specificity by $\frac{TN}{TN+FP}$.

## 6 Results on simulated *C. elegans* data

In this section we provide additional results obtained on *C.elegans* for various read lengths and coverage depths (Table 4). We also performed tests using a 0.5% error rate and obtained similar results (data not shown).

| Corrector | Sensitivity | Specificity | Correction ratio | % Erroneous reads |
|---|---|---|---|---|
| 150-bp reads at 100X coverage ||||
| Bcool | **99.595** | **99.999** | **190.793** | **0.303** |
| BFC | 94.854 | 99.997 | 18.53 | 3.634 |
| Bloocoo | 96.852 | 99.993 | 26.14 | 3.753 |
| Lighter | 97.352 | 99.995 | 31.70 | 2.857 |
| Musket | 98.922 | 99.997 | 73.89 | 1.466 |
| 150-bp reads at 50X coverage ||||
| Bcool | **99.467** | **99.999** | **151.35** | **0.421** |
| BFC | 95.857 | 99.998 | 22.98 | 2.789 |
| Bloocoo | 97.090 | 99.994 | 28.14 | 3.5 |
| Lighter | 98.149 | 99.996 | 43.97 | 1.996 |
| Musket | 98.822 | 99.997 | 68.68 | 1.569 |
| 250-bp reads at 100X coverage ||||
| Bcool | **99.537** | **99.999** | **183.53** | **0.458** |
| Bloocoo | 97.376 | 99.993 | 30.28 | 4.893 |
| BFC | 93.327 | 99.998 | 14.58 | 5.667 |
| Lighter | 97.346 | 99.994 | 31.16 | 4.225 |
| Musket | 99.142 | 99.998 | 90.33 | 1.867 |
| 250-bp reads at 50X coverage ||||
| Bcool | **99.498** | **99.999** | **162.98** | **0.516** |
| Bloocoo | 97.634 | 99.994 | 33.34 | 4.509 |
| BFC | 94.541 | 99.998 | 17.82 | 4.431 |
| Lighter | 98.008 | 99.995 | 40.28 | 3.203 |
| Musket | 99.071 | 99.998 | 84.70 | 1.963 |

Table 4. Correction metrics of various correctors applied on C.elegans reads simulated with a 1% error rate

Additionally, we provide results (Table 5) obtained while using a high $k$ value.

## 7 Results on simulated human chromosome 1 data

In this section we provide additional results obtained on the human chromosome 1 for various read lengths and coverage depths (Table 6). We also performed tests using a 0.5% error rate and obtained similar results (data not shown).

| Corrector | Sensitivity | Specificity | Correction ratio | % Erroneous reads |
|---|---|---|---|---|
| $k = 63$ ||||
| Bcool | **99.395** | **99.998** | **129.504** | **0.519** |
| Bloocoo | 82.933 | 99.992 | 5.590 | 14.210 |
| Lighter | 96.598 | **99.998** | 27.493 | 1.5276 |
| $k = 95$ ||||
| Bcool | **99.590** | **99.999** | **178.904** | **0.321** |
| Bloocoo | 63.5537 | 99.997 | 2.722 | 20.116 |
| Lighter | 61.240 | 99.995 | 2.547 | 25.458 |

Table 5. Simulated C. elegans 150-bp reads with 1% error rate and 100X coverage/ The Musket run was not able to complete and BFC yielded a correction ratio < 1 - both are therefore not reported here.

| Corrector | Sensitivity | Specificity | Correction ratio | % Erroneous reads |
|---|---|---|---|---|
| 150-bp reads at 100X coverage ||||
| Bcool | **99.017** | **99.999** | **95.40** | **0.745** |
| BFC | 86.225 | 99.991 | 6.82 | 10.519 |
| Bloocoo | 92.573 | 99.965 | 9.15 | 9.03 |
| Lighter | 91.269 | 99.975 | 8.96 | 8.709 |
| Musket | 93.052 | 99.982 | 11.4 | 8.307 |
| 150-bp reads at 50X coverage ||||
| Bcool | **98.193** | **99.998** | **48.83** | **1.505** |
| BFC | 87.397 | 99.991 | 7.434 | 9.543 |
| Bloocoo | 93.053 | 99.962 | 9.38 | 8.641 |
| Lighter | 91.915 | 99.976 | 9.550 | 8.123 |
| Musket | 92.876 | 99.982 | 11.19 | 8.458 |
| 250-bp reads at 100X coverage ||||
| Bcool | **99.392** | **100** | **153.73** | **0.577** |
| Bloocoo | 93.291 | 99.964 | 9.76 | 12.308 |
| Lighter | 90.336 | 99.977 | 8.35 | 13.616 |
| Musket | 93.816 | 99.982 | 12.6 | 10.708 |
| BFC | 82.744 | 99.994 | 5.6 | 15.546 |
| 250-bp reads at 50X coverage ||||
| Bcool | **98.855** | **99.999** | **77.376** | **1.214** |
| Bloocoo | 93.774 | 99.962 | 10.038 | 11.798 |
| Lighter | 91.717 | 99.977 | 9.42 | 11.855 |
| BFC | 84.008 | 99.994 | 6.05 | 14.170 |
| Musket | 93.666 | 99.982 | 12.372 | 10.864 |

Table 6. Correction metrics of various correctors applied on human chromosome 1 reads simulated with a 1% error rate

## 8 Results on simulated human chromosome 1 diploid data

Two vcf files were retrieved from the "1000 genome project" (phase 1 release), corresponding to the human chromosome 1 of two individuals: HG00096 and HG00100. We then generated the genome sequences for the two diploids, i.e. two sequences per individual, by placing the substitutions listed in the vcf files onto the human reference sequence (GRCh37/hg19 reference assembly version). A total of 316,502 positions were mutated, with 131,263 positions mutated at the same time in both individuals (representing an average 0.5 SNP per Kb in each individual). 29,038 SNPs (9 %) were homozygous in both individuals (homozygous-homozygous), 218,556 (69 %) were heterozygous in only one individual (homozygous-heterozygous) and the remaining 68,908 (22 %) were heterozygous in both individuals. We then simulated a 100X coverage sequencing with a 1% error rate from this pair of diploid genomes. Results are presented Table 7.

| Corrector | Sensitivity | Specificity | Correction ratio | % Erroneous reads |
|---|---|---|---|---|
| 250-bp reads at 100X coverage | | | | |
| Bcool | **99.506** | **99.998** | **152.60** | **0.685** |
| Bloocoo | 94.045 | 99.966 | 10.71 | 11.415 |
| BFC | 83.996 | 99.994 | 6.04 | 14.256 |
| Musket | 94.314 | 99.982 | 13.461 | 10.298 |
| Lighet | 91.309 | 99.977 | 9.143 | 12.648 |
| 150-bp reads at 100X coverage | | | | |
| Bcool | **99.256** | **99.998** | **100.40** | **0.720** |
| Bloocoo | 93.330 | 99.966 | 9.98 | 8.281 |
| BFC | 87.438 | 99.992 | 7.472 | 9.474 |
| Musket | 93.548 | 99.982 | 12.11 | 7.907 |
| Lighter | 91.414 | 99.977 | 9.176 | 8.667 |

Table 7. Correction metrics of various correctors applied on diploid human chromosome1 reads simulated with 1% error rate

## 9 Results on simulated whole human genome data

In this section we provide additional results obtained on the whole human genome for various read lengths and sequencing depths (Table 8). We also performed tests using a 0.5% error rate and obtained similar results (data not shown).

| Corrector | Sensitivity | Specificity | Correction ratio | % Erroneous reads |
|---|---|---|---|---|
| 150-bp reads at 100X coverage | | | | |
| Bcool | **97.735** | **99.996** | **37.73** | **1.946** |
| Bloocoo | 88.901 | 99.956 | 6.48 | 12.427 |
| Lighter | 85.565 | 99.971 | 5.78 | 13.639 |
| 150-bp reads at 50X coverage | | | | |
| Bcool | **96.495** | **99.996** | **25.66** | **2.916** |
| Bloocoo | 89.591 | 99.953 | 6.63 | 11.962 |
| Lighter | 87.414 | 99.97 | 6.42 | 11.954 |
| 250-bp reads at 100X coverage | | | | |
| Bcool | **98.415** | **99.998** | **56.80** | **1.718** |
| Bloocoo | 89.649 | 99.956 | 6.789 | 16.810 |
| Lighter | 84.969 | 99.973 | 5.65 | 19.639 |
| 250-bp reads at 50X coverage | | | | |
| Bcool | **97.621** | **99.996** | **35.87** | **2.668** |
| Bloocoo | 90.366 | 99.953 | 6.98 | 16.200 |
| Lighter | 87.109 | 99.971 | 6.35 | 16.964 |

Table 8. Correction metrics of various correctors applied to reads simulated from the complete human genome with a 1% error rate

## 10 DBG construction strategies

In this section we evaluate diverse DBG construction strategies. We performed tests on a simulated *C. elegans* 50X data set of 150-bp reads with a 1% error rate. Results are presented in Table 9. We show only graph-cleaning results obtained with low $k$-mer abundance thresholds (2 and 3) as higher values would not make sense for our approach. Results using higher $k$-mer abundance thresholds are showed here for KAF only, as it corresponds to a classical $k$-mer spectrum approach.

| $k$-mer abundance threshold | KAF | KAF+TIP | KAF+ UAF | KAF+TIP+UAF |
|---|---|---|---|---|
| k=31 | | | | |
| 2 | 54,075,339 / 36 | 19,429,473 / 550 | 858,726 / 46 | 642,396 / 560 |
| 3 | 5,628,920 / 55 | 857,469 / 57 | 676,110 / 623 | 676,110 / 623 |
| 4 | 1,968,288 / 75 | | | |
| 5 | 1,115,357 / 85 | | | |
| 10 | 216,094 / 1,175 | | | |
| k=63 | | | | |
| 2 | 31,902,775 / 347 | 1,586,655 / 2,070 | 127,975 / 1,409 | 78,639 / 3,963 |
| 3 | 1,837,789 / 2,507 | 176,693 / 4,612 | 136,972 / 3,295 | 76,407 / 4,653 |
| 4 | 482,095 / 13,145 | | | |
| 5 | 217,508 / 54,225 | | | |
| 10 | 17,508 / 5,083,037 | | | |

Table 9. Evaluation of different DBG construction strategies proposed by Bcool. Each result presents two values ($v_1/v_2$). Value $v_1$ is the number of erroneous $k$-mers present in the DBG and $v_2$ is the number of genomic $k$-mers missing in the DBG. In this experiment (and by default in Bcool) the unitig filtering threshold was set to five. The different rows represent the efficiency of the different strategies tested: $k$-mer abundance filter alone (KAF), tip removal after $k$-mer filter (KAF+TIP), unitig abundance filtering after $k$-mer filtering (KAF+UAF), and the combination of the three strategies (KAF+TIP+UAF).